\documentclass[12pt,a4paper,onecolumn]{article}                
\usepackage{latexsym}
\usepackage[a4paper,left=2.5cm,right=2.5cm,top=2.5cm,bottom=2.5cm]{geometry}
\usepackage{amsmath}
\usepackage{amsfonts}
\usepackage{amsthm}
\usepackage[pagebackref]{hyperref}
\begin{document}

\title{Is the individuality interpretation of quantum theory wrong ?}

\author{U. Klein\thanks{ulf.klein@jku.at}\\University of Linz\\Institute for Theoretical Physics\\ A-4040 Linz, Austria\\}

\date{\today}

\maketitle

\bibliographystyle{plain}

\begin{abstract}

We analyze the question whether or not quantum theory should be 
used to describe single particles. Our final result is that a rational 
basis for such an 'individuality interpretation' does not exist. A 
critical examination of three principles, supporting the individuality 
interpretation, leads to the result that no one of these principles 
seems to be realized in nature. The well-known controversy 
characterized by the names of Einstein (EPR), Bohr and Bell is 
analyzed. EPR proved 'predictive incompleteness' of quantum theory, 
which implies that no individuality interpretation exists. Contrary 
to the common opinion, Bell's proof of 'metaphysical completeness' 
does not invalidate EPR's proof because two crucially different 
meanings of 'completeness' are involved. The failure to distinguish 
between these two meanings is closely related to a fundamentally 
deterministic world view, which dominated the thinking of the 
19th century and determines our thinking even today.

\end{abstract}

%\keywords{Foundations of quantum theory
%\and Individuality Interpretation of Quantum Theory
%\and Statistical Interpretation of Quantum Theory
%}

\section{Introduction}
\label{sec:Introduction}

The question formulated in the title of this essay requires first
clarification of a semantic point. An interpretation cannot be 
wrong in the same sense as an experimentally verifiable theoretical 
prediction. The term 'wrong' is used here in the sense of 'extremely 
misleading'. In this sense a wrong interpretation leads to paradoxical 
contradictions or to internal inconsistencies such as unsolvable 
theoretical problems. I will first formulate some basic assumptions
and explain what 'individuality interpretation' means before I will 
try to answer the question.

I start by formulating the most important assumptions underlying 
this work. A physical theory is a set of equations together with 
a number of rules how to compare theoretical and experimental 
results. Predictions, i.e. real numbers obtained by solving the 
basic equations using input data referring to earlier times, 
represent the core of a physical theory. The interpretation 
of the mathematical terms is also part of a physical theory. 
It gives 'meaning' to the mathematical variables, but does 
not affect the predictive core of the theory. Several different 
interpretations of one mathematical formalism are possible. 
They lead to slightly different physical theories but cannot 
make a physical theory right or wrong, since they do not 
affect the predictive core of the theory. We may say that 
a theory is given by a set of predictions (which constitute the 
invariant core) and an interpretation.

What, exactly, does 'individuality interpretation' mean ? A physical 
theory which allows prediction of single (individual) events, in 
particular predictions about single particles, must necessarily be 
interpreted in this sense. For classical mechanics this individuality 
interpretation is obviously correct. Such a theory could also be called 
a deterministic theory, because it claims to predict the behavior of 
individual particles 'with certainty'. There is no room in such a 
theory for uncertainty, all  predictions of this individualistic 
theory have a probability equal to one . Testing the predictions of 
this theory requires a single experiment. 

On the other hand there are physical theories, expressed in a completely 
'deterministic' mathematical form, whose output cannot be verified 
in single experiments because the scattering of data is too large to 
be neglected. Clearly, the output of a physical theory must be testable. 
If it cannot be tested in individual experiments, essentially the only 
possibility left is a statistical test. In this case the output of 
the theory is given by statistical quantities like probabilities or 
expectation values. These numbers can be compared with 
experimental results and verified or falsified just as the output of the 
deterministic theories discussed above. But in order to do this an 
infinite number of individual systems, all prepared in the same way,
have to be studied experimentally. This is the way statistical 
measurements have to be performed in principle (in practice 
simpler possibilities exist), no other testable meaning (namely 
"frequentist probability') can be ascribed to  the term probability 
in a physical context. A statistical theory can obviously not be used 
o make predictions about individual events, because such predictions 
cannot be verified in individual experiments. 

From the present discussion one would expect - considering the above, 
rather weak assumptions as evident -  that \emph{no} physical theory 
whose output is given by statistical quantities can be interpreted in an 
individualistic sense. In particular, one would expect that this holds true
for quantum theory (QT), whose output is of a probabilistic nature. 
On the other hand, the dominating interpretation of QT 
tells us that quantum mechanics \emph{is} a theory about individual 
particles. We have been using phrases like 'the Schr\"odinger equation of a 
single electron' or 'the quantum mechanical description of 
a single particle' an infinite number of times. 
This kind of talking determines our thinking. The idea that 
QT describes individual events and particles presents the basis 
for much, if not most, of current research on foundations of physics. 

\section[The conceptual basis of the individuality interpretation of quantum theory]{The conceptual basis of the individuality interpretation of quantum theory}
\label{sec:conc-basis-indiv}
According to the assumptions above, an individuality interpretation of 
a probabilistic theory does not make sense. This means that the boundaries 
of this simple framework have to be left far behind in order to establish 
QT as a theory about single particles. There are individuality 
interpretations which require  more than a single universe or 
the participation of the observer's brain~\cite{barrett:quantum}. We 
shall not discuss such interpretations here but restrict ourselves to 
the standard, Copenhagen interpretation (CI).
In order to overcome the fundamental conflict between deterministic 
and probabilistic predictions the CI denies the reality of 
unobserved properties~\cite{mermin:hidden}. The 
properties 'come into being' by the act of measurement in a way which 
is unknown and presents an unsolvable 'measurement problem'. The 
CI 'solves' the fundamental conflict in a sophistic sense because 
it is not the task of a physical theory to make predictions about 
non-existing things. But it does not answer the question how things 
come back to reality. The CI's claim for an individuality 
interpretation may also be expressed by the statement 
that QT is a 'complete' theory as regards the description of 
individual particles; a more detailed analysis of the term 'complete' 
will be given in section~\ref{sec:bohr-einstein-two}.

The CI shows several strange features, which have as a common origin 
the switching forth and back between reality and un-reality 
of properties as observation begins and ends. This problem becomes 
more stringent if two conjugate properties (non-commuting observables) 
have to be measured at the same time. A number of principles 
or concepts have been introduced, by the founders of the CI, in order 
to support the individuality interpretation and to explain its strange 
features. There seem to be essentially three such principles. Let us 
begin with  
\begin{itemize}
\item Heisenberg's uncertainty principle, the well-known inequality 
expressing the impossibility to measure position and momentum of 
a single particle simultaneously with arbitrary high precision.
\end{itemize}
This principle will be referred to as individual uncertainty principle 
(IUP). The IUP presents the most important cornerstone of the CI because 
it supports, if true, the idea that certain (conjugate) properties of 
a single microscopic system cannot be simultaneously real. The second 
concept supporting the CI is
\begin{itemize}
\item the particle-wave duality:  Depending on the experimental situation 
individual microscopic systems may behave either like particles or like 
waves.
\end{itemize}
This idea can be considered as a complement to the IUP. In fact, 
according to the IUP particles have either sharp values of position or 
of momentum, depending on the experimental conditions. The former case 
corresponds to the particle picture, the latter to the wave picture. The 
degree of reality of these two pictures is determined by the measurement 
arrangement. The third idea supporting the CI is expressed as an 
assertion about
\begin{itemize}
\item the classical limit of QT: Classical mechanics may be regarded 
as the limiting case of quantum mechanics when  $\hbar$ tends to 
zero~\cite{dirac:principles_p88}
\end{itemize}
The relevance of this last point for the individuality interpretation 
is obvious. If QT really describes individual particles (for 
nonzero  $\hbar$), then it should not change its character - as 
a theory describing individual particles - if the limit $\hbar \to 0$ 
is performed. This limit must agree with a classical 
individualistic theory, namely classical mechanics. Otherwise 
the CI as an individuality interpretation must be called into 
doubt. 

An important point to note is that no one of these three principles is 
part of the quantum theoretical formalism. This means each one needs 
justification from experiment or theory. A second important point is 
that these principles have been set up in the first half of the last 
century and that enormous technological progress has been made 
since then. A re-examination, taking today's results into account, 
seems useful.
 
\section[Critical discussion of the conceptual basis of the individuality interpretation]{Critical discussion of the conceptual basis of the individuality interpretation}
\label{sec:disc-conc-basis}
The philosophical idea that unobserved things cannot be ascribed 
reality goes back to Aristotle and has Descartes and Berkeley as 
prominent advocates. It is in disagreement with the common sense 
philosophy of physicists but this does, of course, not mean that 
it can be rejected right from the start. Starting our analysis 
with the 
\begin{itemize}
\item Uncertainty principle,
\end{itemize}
we should mention at the very beginning that Kennard's 
inequality~\cite{kennard:quantenmechanik}, which is commonly 
written in exactly the same form as the IUP, 
has a very different meaning. It is a statistical relation which has 
nothing to do with the simultaneous measurement of position and 
momentum but expresses the relationship of the statistical fluctuations
of independently measured quantities. Its mathematical derivation from 
the framework of QT is still sometimes erroneously considered as a
confirmation of the IUP. The second point to note is that Heisenberg's
famous Gedanken experiment connects the error in position to the 
disturbance in momentum, and not to the error in momentum (the momentum 
is assumed to be accurately known at the time of position 
measurement); see e.g. Margenau and Park~\cite{margenau:physics_semantics}.  
Therefore it seems that the universal acceptance of the IUP is based on 
two historically-grown misunderstandings, namely the failures to 
distinguish individual from statistical measurements and 
measurements from preparations. 

Considerable efforts have been undertaken to derive 
Heisenberg-like relations from QT. This requires, however, 
additional assumptions, forcing individuality concepts into
the statistical formalism of QT. The most important of these is 
the 'projection postulate', the assumption that the state vector 
jumps after a measurement into an eigenspace of the corresponding 
operator. The details of this process, which is sometimes referred 
to as 'non-unitary time evolution', are unknown. This beautiful
expression is used to hide the fundamental difference between 
individual and statistical predictions. The projection postulate 
was suggested by the ingenious mathematician von Neumann and 
seems very convincing from the point of view of mathematical 
simplicity. But it may be a simple error. In this context it 
must be mentioned that von Neumann's proof of the non-existence 
of hidden variables contains a simple error. It was quoted 
many times, as an argument in favor of 'completeness' 
of QT, during a period of more than thirty years (!), until 
the 'silly error' in its derivation became widely 
known~\cite{mermin:hidden}.  All attempts to derive 
Heisenberg-like relations from QT use this projection postulate, 
as well as other assumptions formulated in the abstract 
language of Hilbert space~\cite{hofmann:measurement}. Depending on 
the chosen assumptions some authors derive  relations similar to 
Heisenberg's inequality~\cite{busch:uncertainty} while others obtain 
different expressions~\cite{ozawa:universally}. Several 
experimental violations of Heisenberg's inequality have been reported, 
the most recent one by Erhart et al.~\cite{erhart_et_al:experimental}. 

Most relevant for the CI's claim of un-reality of unobserved 
properties is the IUP, i.e. Heisenberg's inequality  interpreted 
as a relation between measurement errors of conjugate properties. 
Astonishingly, the practical basis for the IUP seems to be still 
Heisenberg's famous light-microscope Gedanken experiment - despite 
the fact that it says nothing about the simultaneous measurement of 
position and momentum. Thus, let us  first ask if other 
"Gedanken experimente" have been designed which show a violation of 
the IUP. This is indeed the case. Such idealized measurement 
arrangements have been proposed 
by Prugovecki~\cite{prugovecki:measurement}, Park and 
Margenau~\cite{park:simultaneous}, 
Ballentine~\cite{ballentine:statistical}, 
Popper~\cite{popper:quantum_schism} and others. We may, secondly, 
ask, as a question of primary importance, if the IUP has ever be 
confirmed experimentally. Not a single experimental confirmation 
has been reported~\cite{busch:uncertainty} since Heisenberg's creation of 
the IUP in 1927. On the other hand, data showing violations 
of the IUP have been published. We mention, in particular, the 
realization of Popper’s thought 
experiment~\cite{popper:quantum_schism} by Kim and Shih~\cite{kim_shih:realization_popper}. 

Summing up, we find no experimental or theoretical facts supporting 
the IUP. This principle does not seem to be an element of science, 
but rather a historically-grown habit or an object of quasi-religious 
admiration.
\begin{itemize}
\item Particle-wave duality
\end{itemize}
Recent experiments by Tonomura~\cite{tonomura:demonstration} 
and others have shown that single particles are always particles and 
never waves. A 
\href{http://www.hitachi.com/rd/research/em/movie.html}{video} on 
the Hitachi website~\cite{tonomura:double_slit_movie} shows the 
development of a double-slit interference pattern as a consequence 
of an increasing number of electrons arriving at the screen. As 
pointed out by Silverman in his discussion of the 
Tonomura experiment~\cite{silverman:quantum}:
"The manifestations of wave-like behavior are statistical in 
nature and always emerge from the collective outcome of many 
electron events" Thus, no mysterious transformation between particles 
and waves is required. The origin of the miraculous 
'particle-wave duality' is poor resolution of early experimental data.
\begin{itemize}
\item The classical limit
\end{itemize}
The idea that classical mechanics must emerge as the classical limit of 
QT was advocated by Bohr, Dirac and others. But this idea led to a 
large number of open questions and contradictions. The problem becomes 
much simpler if one admits the possibility that the classical limit of 
QT differs from classical mechanics. It has been mentioned before 
that a straightforward application of the limit $\hbar \to 0$ to 
Schr\"odinger's equation leads to a classical probabilistic theory and 
not to classical 
mechanics~\cite{ballentine:quantum,klein:schroedingers,klein:website}  
A recent, more complete treatment~\cite{klein:what} leads to the same
conclusion. 

To summarize, closer examination shows that neither 
the IUP, nor the wave-particle duality, nor the claim that classical 
mechanics emerges as the classical limit of QT present physically 
well-defined concepts. No support is provided for the philosophical 
idea that unobserved properties are not real and for the related 
idea that an individuality interpretation of QT exists. On top of 
that, this also implies that a fundamental and very successful 
methodical principle of physics, namely the principle of reductionism, 
can not be universally valid. This principle is not compatible 
with the statistical interpretation of QT. As is well-known, the 
scientific community decided to keep the philosophical dogma of 
reductionism along with the individuality interpretation of QT. 
From a psychological point of view this is understandable, since 
we expect science to yield predictions with certainty, but the 
question is how much weight should be given to psychological 
expectations.
 
\section[EPR, Bohr, Bell, and two meanings of 'completeness' ?]{EPR, Bohr, Bell, 
and two meanings of 'completeness'}
\label{sec:bohr-einstein-two}
Present research on foundations of QT is strongly influenced by
a paper published in 1935 by Einstein, Podolsky, and Rosen 
(EPR)~\cite{einstein.podolsky.ea:can}. There is an enormous secondary 
literature, see e.g. Fine~\cite{fine:shaky}, 
Ballentine~\cite{ballentine:statistical}, 
Redhead~\cite{redhead:incompleteness}, and the author's   
\href{http://statintquant.net}{website}~\cite{klein:website}. 
In this work, EPR claim that the quantum-mechanical description 
of reality is incomplete. The CI is attacked 'from inside' because 
the basic assumptions used in this paper do not reflect the 
positions of the authors but are part of the CI. The most 
significant example is EPR's assumption that "The state of the 
particle is completely characterized by a wave function $\psi$", 
a statement in sharp opposition to Einstein's well-documented 
opinion that $\psi$ describes an ensemble. EPR's conclusion was, 
of course, attacked by CI's advocates who considered QT as a 
complete theory. However, a discussion of the specific EPR problem 
was generally avoided and EPR's claim of incompleteness of QT 
was attacked on different routes - circumventing the specific 
problem. Bohr, in his reply, took a very philosophical, elusive 
route, which was not really convincing for many people.

According to the prevailing opinion this question was decided in favor of
Bohr by the work of John Bell~\cite{bell:on,bell:on*1}, about 
thirty years later. Bell circumvented the specific EPR problem by 
relating it to the problem of hidden variables. A (local) hidden 
variable theory is compatible with all predictions of QT providing, 
however, at the same time, a more detailed (deterministic) description 
of reality. Physical intuition tells us that such a thing cannot exist 
but Bell proved that it cannot exist - at least within the framework 
of his postulates; all no-go proofs are of course only valid within a 
certain 'universe of discourse' (repeated remarks on this important 
limitation will be omitted from now on for brevity). He formulated 
general conditions for local hidden variable theories  and derived 
therefrom an inequality which \emph{differs} from the corresponding 
prediction of QT. Thus, he showed that hidden variable theories cannot 
exist if QT is correct. This shows that QT is a 'complete' theory 
(with the meaning of 'complete' given in context).  
This reasoning seems correct but the question is what can be concluded 
from it. The simplest conclusion is that EPR's proof of incompleteness 
of QT cannot be true because Bell showed that QT is complete. I claim 
that this simple reasoning is not justified because a subtle semantic 
trap, concerning the meaning(s) of the word 'complete', has been 
overlooked. 

The word complete has two different meanings. If used to 
characterize the predictive power of a physical theory it means:
"All facts that can be observed can be predicted (with certainty)''.
This kind of completeness could be called 'predictive completeness', 
or 'p-completeness' for brevity. In order to find out if a physical 
theory is p-complete one needs solutions (predictions) of this theory 
and experiments testing these predictions. This first kind of 
completeness may equivalently be characterized by saying that an 
'individuality interpretation' for this (p-complete) theory exists. The 
standard example for a p-complete theory is classical mechanics. 
Classical massless field theories are of a similar nature but do 
not directly describe individual particles.

The second meaning of the word complete can be described as 
follows: "No better theory, in the sense of producing more 'definite' 
(deterministic) predictions, exists". This is a very strong assertion. It 
entails not only the concrete physical theory under discussion but 
also an infinite number of other theories (all unknown),  which are 
all not allowed to exist according to the assertion. Such an assertion 
can of course only be verified within a certain 'universe of discourse', 
which may possibly be generalized in later steps. But it can be approached 
nevertheless. Let us call this second kind of completeness 'metaphysical 
completeness', or 'm-completeness'. As an example, we mention 
classical probabilistic theories where the uncertainty is only in the 
initial conditions while the movement in phase space is 
deterministic~\cite{klein:statistical}. These theories are  
m-incomplete, with classical mechanics playing the role 
of the 'better theory'.

How are these two kinds of completeness related to each other ? 
A p-complete theory is also m-complete. It would not make sense 
to search for a better theory than classical mechanics (in its 
range of validity) because classical mechanics makes already 
predictions with probability equal to one. This means, 
the implication 
\begin{equation}
  \label{eq:alogicalimplication}
\mbox{p-completeness }\Rightarrow\mbox{m-completeness}
\mbox{,}
\end{equation}
holds true. This means that m-incompleteness implies 
p-incompleteness and that p-incompleteness is 
a necessary condition for m-incompleteness. On the 
other hand  p-incompleteness is not a sufficient condition 
for m-incompleteness. A p-incomplete theory may 
be either m-complete or m-incomplete.

Let us now reconsider the EPR-Bohr-Bell question using this refined
vocabulary. What Bell proved is obviously m-completeness 
of QT. In EPR's paper  both kinds of completeness occur. In the last 
paragraph EPR express their believe that QT is m-incomplete:
\begin{quote}
"We believe however that such a [more complete] theory is possible"  
\end{quote}
The communication of EPR's 'believe' (which had been known for 
a long time) is of course not the central message of EPR. The central 
message is given by the logical deductions reported in the body of the 
paper, i.e. in the whole paper except the last paragraph. So, which 
kind of completeness is referred to in the body of EPR's paper ? The 
subject of the paper is the problem of predictions of the values 
of certain observables, thus what EPR mean by completeness is obviously 
p-completeness. An assertion of m-incompleteness, i.e. a statement 
that a better theory then QT must exist, can nowhere be found in 
the relevant part (the body) of EPR's paper. 

The proof of p-incompleteness of QT was of course a necessary 
prerequisite for EPR's 'believe' in a deterministic replacement 
of QT. If an analysis had led to the conclusion that QT is p-complete, 
this had also implied that QT is m-complete. On the other hand, EPR 
were aware of the fact that p-incompleteness is only a necessary and 
not a sufficient condition for m-incompleteness. Thus, they were aware 
of the fact that their proof of p-incompleteness of QT did not imply 
m-incompleteness of QT. They express this in the first sentence of the 
last paragraph of their paper in a very clear way:  
\begin{quote}
"While we have thus shown that the wave function does not provide a 
complete description of physical reality, we left open the question 
of whether or not such a description exists".
\end{quote}
It is a real mystery why this clear statement, separating cleanly the 
two different kinds of completeness (made even more explicit in the 
present essay only by means of different names) from each other, has 
been overlooked by the scientific community. 

It follows that Bell's proof of m-completeness cannot be used, even 
if we accept the assumptions underlying his 
proof~\cite{hess_et_al:hidden}, as an argument against EPR's 
proof of p-incompleteness of QT.  Bell was in error, when claiming 
that his results contradict EPR ! If we accept both Bell's and EPR's 
findings we arrive at the final conclusion that QT is p-incomplete and 
m-complete. This conclusion is compatible with a recent derivation 
of non-relativistic quantum theory from statistical 
postulates~\cite{klein:statistical,klein:nonrelativistic} 
and  agrees roughly with the common sense assessment of QT as a 
correct (complete) statistical (incomplete) theory.  It implies 
that an individuality interpretation of QT is not justified. 

The present conclusion can only be avoided if one takes a deterministic 
point of view of the world, namely that everything that can be observed 
must in principle be predictable. This then means that p-incompleteness 
implies m-incompleteness (or the existence of a hidden variable theory). 
But note that this implication, which eliminates our final conclusion, 
is not a logical requirement but the consequence of a new (in fact, very 
old) philosophical dogma. From the point of view of physics such a 
deterministic dogma is not required. Interestingly, this deterministic 
point of view was \emph{shared} by Einstein and 
Bohr (see~\cite{klein:website} for a more detailed explanation), despite  
their otherwise very different opinions. It was denied by several other 
authors, in particular by Popper~\cite{popper:open_universe}. 
Unfortunately, today's discussions are  still centered around the 
two alternatives represented by Einstein and Bohr. 

\section[More recent developments]{More recent developments}
\label{sec:more-recent-devel}

Recent years have shown a tendency to reinterpret the body of EPR's 
paper as a proof of m-incompleteness. The sharp contrast between 
the 'believe' in the last paragraph and the science in the body of 
the paper had to be hidden somehow. The paper was so to say 'rewritten' 
as a hidden variable theory, 'elements of reality' became hidden variables. 
However, the real EPR paper is not a hidden variable theory but 
the construction of  an internal contradiction within the CI. It is 
certainly true that EPR's conclusion implies the \emph{possibility} 
of a more complete description. However they have not \emph{shown} 
that such a more complete description exists. The only way to do 
this is to construct a 'better' (hidden variable) theory for QT. 
Nothing like that can be found in EPR's paper. 

As a further development, which will be dealt with only briefly, the 
validity of EPR's final conclusion has been discussed in conjunction 
with the validity of their basic assumptions. According to the late 
Einstein, EPR's final conclusion may be rewritten as the statement that 
not both of the following two assertions (quoted literally 
from~\cite{einstein:reply}) can be true:
\begin{enumerate}
\item The description by means of the Psi-function is \emph{complete}.
\item The real states of spatially separated objects are independent from each other.
\end{enumerate}
Today, the second assertion is frequently split into the 
assertions of 'locality' and 'reality' (see e.g. 
Wiseman~\cite{wiseman:from_einsteins}). Incompleteness with regard to 
predictions of individual events is a familiar feature. On the 
other hand a breakdown of 'locality' or 'reality' presents a much 
stronger and stranger assumption. Thus, the failure of the 
first assertion seems more natural. Therefrom Einstein's  
conclusion that QT is 'incomplete'. 

If Bell's theorem is (erroneously) used to show that the first assertion 
is true, then the breakdown of at least one of the fundamental scientific 
principles of 'locality' or 'reality' is a necessary consequence. Thus, 
certain strange features associated with single events became a subject 
of intense research because they could now be described in terms of 
c.f. a 'breakdown of locality'. Thus the strange ('weird', 'magical') features 
of QT which always appear, whenever QT is used  to describe single 
particles, became manifest once again, but this time in a supposedly 
more definite form thanks to Bell's theorem. 

If, on the other hand, the two different meanings of the term 
'completeness' are clearly distinguished from each other, then Bell's 
proof of m-completeness cannot be used to eliminate QT's property 
of p-incompleteness. Then, the above Einstein alternative leads to the 
same final conclusion as before, namely that an individuality 
interpretation of QT does not exist. This means that QT is a statistical 
theory (probably complete in a metaphysical sense, and certainly 
complete in a metaphysical sense as defined by Bell) which by its 
very nature cannot be used to describe the behavior of single particles. 
The 'strangeness' of QT is nothing but the consequence of an unjustified 
extension of its range of validity. Typically, all the strange things 
never happen in the laboratory but always in the brain of the interpreter.

\section[Conclusion]{Conclusion}
\label{sec:conclusion}
In the first part of this essay we found that a rational basis 
for several principles, believed to support the philosophical 
basis of the CI, does not exist. The mysterious nature of the CI, 
which contains undefined statements in its definition, can be 
expected to lead to further obscurities whenever applied to single 
particles. This is indeed what happens. An attempt by EPR to make 
the contradictory character of the CI explicit, i.e. to prove the 
p-incompleteness of QT - or the non-existence of an individuality 
interpretation for QT - was successful. Bell's proof of m-completeness 
cannot be used to invalidate EPR's proof of p-incompleteness. Our 
final conclusion is that QT is p-incomplete and m-complete. The 
common failure to distinguish the two different meanings of completeness 
is due to an old 'deterministic dogma' which rules our thinking even 
today. This dogma invalidates, if true, our final conclusion. 
However, this deterministic point of view is not a logical neccessity 
but rather a historical grown intellectual habit. According to the 
present analysis it is incompatible with the structure of QT 
and should be abandoned.

\bibliography{uftbig}

\end{document}